\documentclass[preprint,a4paper,11pt]{elsarticle}
\pdfoutput=1

\usepackage[english]{babel}

\journal{{\tt arXiv}}

\begin{document}

\begin{frontmatter}
\title{How to choose a PIN -- assessment of dictionary methods}

\author[stu]{{L\raise.05ex\hbox{\kern-.30em '}\kern0.05em}ubica Stanekov\'{a}}
\ead{ls@math.sk}
\author[uk]{Martin Stanek}
\ead{stanek@dcs.fmph.uniba.sk}

\address[stu]{Department of Mathematics and Descriptive Geometry, 
  Faculty of Civil Engineering, Slovak University of Technology, Radlinsk\'{e}ho 11,
  813 68 Bratislava, Slovakia}
\address[uk]{Department of Computer Science, Faculty of Mathematics, Physics 
  and Informatics, Comenius University, Mlynsk\'{a} dolina, 842 48 Bratislava, Slovakia}
\date{\normalsize \today}

\begin{abstract}
Personal Identification Numbers (PINs) are commonly used as an authentication mechanism. 
An important security requirement is that PINs should be hard to guess for an attacker. 
On the other hand, remembering several random PINs can be difficult task for an individual. 
We evaluate several dictionary-based methods of choosing a PIN. We experimentally show 
that these methods are far from ideal with respect to expected covering of the PIN space 
and the entropy of PINs. We also discuss two methods for constructing easy to memorize
PIN words for randomly chosen PINs.
\end{abstract}


\end{frontmatter}

\section{Introduction}

Various forms of authentication are used in our everyday life. Some of them employ a PIN, 
i.e. a string of digits with fixed length,  usually between 4 and 6. 
PINs are used in various applications: for cardholder verification in financial services \cite{EMV},
for authorized personnel verification in physical access control systems, for SIM (Subscriber 
Identity Module) card owner verification in mobile phones, etc. The ubiquitous nature of PINs and the 
fact that a simple numpad is often the only practical option in various environments mean that PINs stay 
here as a viable authentication method for a long time. 

The security of PIN based authentication depends on multiple factors and subjects involved in a particular
application. A comprehensive description of issues in PIN management throughout its entire life cycle
can be found in ISO standard 9564 \cite{ISO} or standards proposed by PCI Security Standards Council,
such as \cite{PCI}. Practically oriented guidelines on PIN security can be found in \cite{VISA}. A user 
of an application usually controls only his/her own PIN and not other security measures. Therefore 
there are some common recommendations for users, such as never share your PIN with anyone, 
memorize your PIN (do not write it down), select a PIN 
that cannot be easily guessed, be aware of shoulder surfing, etc. 

Recently published study on frequency of 4-digit PINs \cite{PINstudy} reveals that most
people fail at the one of the basic recommendation: your PIN should be unguessable. However
the recommendations give almost no hints on how to actually choose a PIN. They say how 
\emph{not} to select a PIN, e.g. using sequential digits such as 1234 (ranking as the most frequent PIN 
in the cited study), repetitive digits such as 1111 or 0000 (five spots in top 10), birthdays or anniversaries 
(each 19?? combination can be found in the top 20\% of the dataset in the cited study), etc. 
In an ideal world everyone would use and remember randomly and uniformly generated PINs of required 
length. Nevertheless,  our ability to memorize random PINs is limited, even more so when an individual 
must remember multiple PINs for distinct applications.

The problem can be approached from two different sides:
\begin{enumerate}
\item Methods of generating easy to remember PINs, usually with some alpha-numeric aids (e.g. various tips
  of this kind can be found  in \cite{Pri}). The advantage of this approach lies in easy to apply techniques.
  On the contrary, it is usually difficult to assess the entropy of PINs constructed this way.
  A variant of this approach is using PINs derived from passwords \cite{JL11,JL13}.
\item Methods to memorize randomly chosen PINs. This ensures highest entropy of used PINs.
  On the other hand, an individual must develop or learn some memorization technique 
  in order to deal with large number of PINs. For example, there  are several known techniques 
  usually aimed at remembering a long sequences of digits, such as Major 
  System or Dominic System. These techniques are also useful in situations when a user is not allowed to 
  change the PIN.
\end{enumerate}

Low entropy of a PIN yields higher success probability of the guessing attack. A common security measure is to 
limit the number of unsuccessful authentication attempts, usually to three, and block the access or the 
device after reaching this number. Even with such measure in place, users should use as unpredictable PINs as
possible.

\paragraph{Our contribution and results} 
In this paper we focus on the problem of choosing a PIN. Formally, a PIN is an element 
from the set $\{0,1,\dots,9\}^n$, for an integer $n$. The value $n$ is the \emph{PIN length} and we use the most common 
values $n=4$ and $n=5$ in our experiments. A \emph{PIN word} is a string of alphabetic characters of length $n$ that  
is translated by some mapping into a PIN. We mostly use the common mapping from letters to numbers 
offered by keyboards of ATMs, Point-of-Sale (PoS) terminals or mobile phones. Other mappings can be easily 
employed in applications by using pictures attached to an authentication dialog or by customizing the keyboard of 
an authentication device.

The contribution of the paper can be summarized as follows.
\begin{enumerate}
\item We experimentally estimate the entropy and the coverage of the PIN space for dictionary PINs, i.e. PINs 
  obtained from PIN words of the corresponding length from a dictionary (Section \ref{sec-stats}). We analyze 
  dictionaries of various sizes and languages (English, French, German and Slovak). Even though the results 
  are significantly influenced by the size of a dictionary, obtained entropies and PIN space coverings are
  unsatisfactory in comparison to ideal values. 
  On the other hand, the results are comparable with the method of deriving PINs from passwords 
  \cite{JL11,JL13}. In addition, using dictionary PINs is certainly better than using birthday or some of the 
  most frequent PINs (e.g. there are at most 5 dictionary PINs in the top 20 most frequent PINs from \cite{PINstudy}).
\item We test various natural ideas to improve the covering and the entropy of dictionary PINs in Section 
  \ref{sec-improve}. The experiments show mostly negligible improvements for individual ideas.
\item Another approaches are explored in Section \ref{sec-hmm}: translation of randomly generated PINs to
  PIN words or PIN phrases with the aim that these will be easier to remember. Our first proposal is the 
  construction of PIN words using the hidden Markov model of the particular language. The second approach is to
  construct PIN phrases from PINs. 
\end{enumerate}

\section{Statistics of Dictionary PIN Codes}
\label{sec-stats}

In this section we analyze the statistics of dictionary PINs. We use the spell-checking dictionaries from 
LibreOffice suite as a source of PIN words in our experiments. We analyze four 
languages: English (US dictionary, ver. 3.0), French (modern dictionary, ver. 4.8), German (frami dictionary, 
ver. 2012.06.17) and Slovak (ver. 2011.02.28). All dictionaries use supplemental affix files to facilitate 
spell-checking. However, we took our input just from the dictionary (dic) files alone. The dictionary files differ 
significantly in number of words: 

\begin{center}
\begin{tabular}{lr}
  Dictionary & \#words \\
  \noalign{\smallskip}\hline\noalign{\smallskip}
  English & 797\,865 \\
  French  & 72\,474 \\
  German  & 220\,030 \\
  Slovak  & 264\,944
\end{tabular}
\end{center}

One can certainly use other dictionaries, or individual's vocabulary can contain words not available 
in a spell-checking dictionary, or one can combine multiple dictionaries (knowing foreign languages), etc. 
On the other hand, the size of common person's vocabulary is usually much smaller than the number of
words in considered dictionary files \cite{Vocab91,Vocab95}.

In order to allow a translation of the particular word into a PIN we transform each word according to
the following rules: 
\begin{enumerate}
\item remove diacritics from all letters (e.g. \v{s} $\mapsto$ s, \"{o} $\mapsto$ o, \c{c} $\mapsto$ c etc.),
\item strip all non-alphabetic characters (e.g. remove apostrophes etc.)
\item convert word to lowercase.
\end{enumerate}

\noindent Let us note that words that differ only in accented letters were counted as distinct words
(e.g. ``mole'' and ``m\^{o}le'' in French dictionary), but words that differ only in the case of their letters
were counted as the same word (e.g. ``Amos'' and ``amos'' in English dictionary). There were also some special treatments 
of particular dictionary files. For example, the ending 's was striped from all words 
in English dictionary. Moreover, words with multiple occurrences were counted just once. 

The experiments use a standard (the most common) mapping of English alphabet into digits:

\begin{center}
\begin{tabular}{l@{$\;\mapsto\;$}l@{\hskip4ex}l@{$\;\mapsto\;$}l}
  a,b,c & 2  &  m,n,o & 6 \\
  d,e,f & 3  &  p,q,r,s & 7 \\
  g,h,i & 4  &  t,u,v & 8 \\
  j,k,l & 5  &  w,x,y,z & 9
\end{tabular}
\end{center}

Obviously, omission of digits 0 and 1 means that it is impossible to cover all PINs. Comparison of
basic facts for PINs (with length 4 or 5) using 10 vs. 8 digits alphabet is summarized in Table \ref{tab-facts}. 
The entropy is calculated as Shannon entropy (in bits), assuming that the possible
PINs are distributed uniformly.

\begin{table}[h]
\begin{center}
\begin{tabular}{crrrrr}
  & \multicolumn{2}{c}{All digits} & \multicolumn{2}{c}{8 digits} \\
  PIN length &  \#PINs  &  Entropy &  \#PINs  &  Entropy \\
  \noalign{\smallskip}\hline\noalign{\smallskip}
     4  &  10\,000 & 13.29 &  4\,096 & 12.00 \\
     5  & 100\,000 & 16.61 & 32\,768 & 15.00 \\
  \noalign{\smallskip}\hline
\end{tabular}
\caption{Number of possible PINs and the corresponding entropy}
\label{tab-facts}
\end{center}
\end{table}

Besides using words not in dictionaries, one can also use different and more complicated rules to transform 
words into PINs. For example occasional/randomly chosen translations like l/i $\mapsto$ 1, o $\mapsto$ 0
can cover other PINs and potentially increase the entropy. Even more complicated rules can be imagined 
for alphabets with diacritics (e.g. adding/subtracting 1 from digit value when the character contains 
a diacritical mark) -- the possibilities are almost endless. Some ideas are evaluated in Section \ref{sec-improve}.
However, we try to keep things as simple as possible here and we believe that this approach provides good 
estimates for the entropy of dictionary PINs.

The results obtained for particular dictionaries and PINs of the length 4 and 5 are presented in Table 
\ref{tab-results}. The table shows the number of PIN words (i.e. dictionary words of the required length 
translated into a PIN), the number of unique PINs obtained from these PIN words, the covering 
(as a percentage of all possible PINs rounded to an integer value) and the entropy of such PINs. 
The entropy is calculated from frequencies of particular PINs.

\newcommand{\CA}{\multicolumn{1}{c}{4}}
\newcommand{\CB}{\multicolumn{1}{c}{5}}
\begin{table}[h]
\begin{center}
\begin{tabular}{lrrrrrrrr}
  & \multicolumn{2}{c}{English} 
  & \multicolumn{2}{c}{French}
  & \multicolumn{2}{c}{German}
  & \multicolumn{2}{c}{Slovak} \\
  PIN length & \CA & \CB& \CA & \CB& \CA & \CB& \CA & \CB\\
  \noalign{\smallskip}\hline\noalign{\smallskip}
  \#PIN words & 10\,484 & 25\,104 & 1\,571 & 3\,595 & 1\,982 & 3\,562 & 4\,537 & 10\,452\\
  \#PINs &  3\,101 & 12\,979 & 1\,073 & 2\,757 & 1\,334 & 3\,005 & 2\,225 & 7\,133\\
  \noalign{\smallskip}\hline\noalign{\smallskip}
  Covering     &  31\% &  13\% & 10\% &   3\% &  13\% &   3\% &  22\% & 7\% \\
  Entropy      & 11.28 & 13.37 & 9.86 & 11.28 & 10.20 & 11.46 & 10.82 & 12.60 \\
  \noalign{\smallskip}\hline
\end{tabular}
\caption{Statistics for dictionary PINs}
\label{tab-results}
\end{center}
\end{table}

The table clearly shows that the results are influenced by the size of the particular dictionary.
However, the differences in entropies are not as big as one would expect considering the differences
in the number of PIN words (e.g. compare results for English and French dictionaries). It seems 
that after some threshold, extending dictionary/vocabulary adds only little to the overall entropy
of resulting PINs. There is also a large gap between theoretical maximums from Table 
\ref{tab-facts} and values shown in Table \ref{tab-results}. Moreover, it is substantially harder
to cover the PIN space with increasing PIN length. We observe covering only 13\% of 
the entire space of possible PINs, for the PIN length 5 and the largest dictionary. 
Therefore, we focus on possible remedies in the following sections.

Interesting observations can be made by looking at the most frequent PINs. As expected,
the PINs reflect unbalanced distribution of letters in a language, combined with distribution
of opening/closing letters in dictionary words. Therefore, the most frequent PINs usually
show some patterns, especially for the first and the last digits. For example, the digit 9 (besides expected
digits 0 and 1) is missing in the most frequent PINs in all languages, thanks to relatively
low frequency of letters w, x, y, and z. For all dictionaries, we present the most frequent PINs 
obtained from analyzed dictionaries together with examples of the corresponding PIN words in the following 
list:
\begin{description}
\item English \\
  Length 4: The most frequent PINs are 2667 (for words such as ``amor'', ``amos'',
  ``cons'', etc.), 5277, 7377, 7467 and 7667, each one occurring 15 times. With one exception, each
  PIN with frequency greater than 13 ends with digit 7. Moreover, digits 7 and 6 are the most 
  frequent digits among these PINs. 
  \smallskip \\
  Length 5: The most frequent PINs are 72737 (``paper'', ``pards'', ``raper'', etc.), 
  72937, 76737, 46637 and 22737, each one occurring 14 times.
\item French \\
  Length 4: The single most frequent PIN, with 9 occurrences, is 7243 (``page'', 
  ``paie'', ``sage'', ``saie'', etc.). The most frequent PINs end with digit 3,
  caused by words mostly ending with letter `e'.
  \smallskip \\
  Length 5: Having 7 occurrences, the most frequent PINs are 66473 (``Moire'', 
  ``mo\"{i}se'', ``omise'', etc.) and 27383. Similarly to the previous case, the usual end digit is 3.  
\item German \\
  Length 4: The most frequent PINs are 5246 (``jahn'', ``kain'', ``lahm'', etc.) 
  and 5346, both with 7 occurrences.
  \smallskip \\
  Length 5: The most frequent PINs are 62436 (``magen'', ``m\"{a}hen'', ``nagen'', etc.) 
  and 62437, both with 6 occurrences. The usual end digits are 3, 6 and 7 for both PIN lengths.
\item Slovak \\
  Length 4: There is only one PIN with 11 occurrences: 7824 
  (``puch'', ``s\'{u}ci'', ``r\'{u}\v{c}i'', etc.). Eight most probable PINs start with
  digit 7.
  \smallskip \\
  Length 5: With frequency 9, the PIN 78652 (``stoja'', ``stoka'', ``\v{s}t\'{o}la'', 
  etc.) is the most probable. The most frequest PINs start with digits 7 or 8 and end with 
  digits 2 or 8.
\end{description}

An interesting result is the comparison of dictionary PINs with top 20 most frequent PINs
from the study \cite{PINstudy}. Let us note that the study analyzed the PINs with length 4, and
according the study almost 27\% of all PINs can be guessed by testing PINs from the top 20.
There are only 5 dictionary PINs that can be found in the top 20 for English dictionary (4444, 2222, 
3333, 6666, 8888). For French, German and Slovak dictionaries we get 4, 3, and 5 dictionary PINs 
in the top 20, respectively. Similarly to English, all such dictionary PINs are repetition of four 
identical digits. Small number of dictionary PINs in the top 20 can be partially explained by the fact
that 11 PINs from the top 20 contain digits 0 or 1.

\section{Improvements}
\label{sec-improve}

The observed covering of dictionary PINs ranges, mostly depending on the dictionary size, from 11\% 
to 31\% for the PIN length 4, and from 3\% to 13\% for the PIN length 5 (see Table \ref{tab-results}). 
In this section we evaluate 
some ideas to improve this situation. Let us emphasize that these ideas are not a complete
list of what can be done. One can certainly come up with other methods to increase the PIN space 
covering as well as the entropy of resulting PINs. This section illustrates what results
can be expected with some natural ideas.

\paragraph{A polyglot method} This method draws a PIN word uniformly from the union of multiple
dictionaries. Of course, there are few people having such vocabulary. On the other hand, with 
some help of a software that offers user a randomly chosen PIN word, this method can be potentially
used by everyone (with additional benefit of learning new foreign words). We decided to combine
all dictionaries (English, French, German and Slovak) in our experiment. Certainly, obtained 
PINs cannot cover more than the sum of the coverings for particular dictionaries, e.g. 26\% 
for the PIN length 5. The results are rather disappointing -- the following table shows only
a moderate improvement for PIN lengths 4 and 5 in comparison to the single dictionary method with 
the largest dictionary (English dictionary in our experiments).

\begin{table}[h]
\begin{center}
\begin{tabular}{lrr}
  PIN length & \CA & \CB \\
  \noalign{\smallskip}\hline\noalign{\smallskip}
  \#PIN words & 18\,574 & 42\,713 \\
  \#PINs &  3\,476 & 16\,568 \\
  \noalign{\smallskip}\hline\noalign{\smallskip}
  Covering     & 35\% & 17\% \\
  Entropy      & 11.38 & 13.63 \\
  \noalign{\smallskip}\hline
\end{tabular}
\caption{Results for polyglot method}
\label{tab-polyglot}
\end{center}
\end{table}

\paragraph{Translations covering digits 0 and 1} This idea was briefly mentioned in Section 
\ref{sec-stats}. In order to cover unused digits 0 and 1 we extend the mapping from letters to 
digits. We choose to map letters `l' and `i' to 1 and letters `o' and `z' to 0; these mappings
were chosen as easy to remember. This simple method provides only a slight improvement of  
coverings for particular dictionaries and PIN lengths. The biggest change was the increase from
31\% to 46\% for English dictionary and the PIN length 4. The improvement was rather negligible for 
all other combinations of dictionaries and PIN lengths -- after rounding to integers, the 
covering percentages for the PIN length 5 changed from 13\% to 16\% for English dictionary, and 
remained unchanged for French, German and Slovak dictionaries.

\paragraph{Change letter mapping} The easiest way to cover all digits is to stretch a mapping
from alphabet to digits. In order to preserve usability, the consecutive sets of letters should be 
mapped to consecutive digits, considering the shape of particular numpads. An example of possible
mapping for a common numpad shape is shown in Figure \ref{fig-stretch}.
\begin{figure}[h]
\begin{center}
\begin{tabular}{l@{\hskip1ex}l@{\hskip4ex}l@{\hskip1ex}l@{$\;\hskip4ex\;$}l@{\hskip1ex}l}
  a,b   & $\mapsto 1$  &  c,d  & $\mapsto 2$ &  e,f   & $\mapsto 3$ \\
  g,h,i & $\mapsto 4$  & j,k,l & $\mapsto 5$ &  m,n   & $\mapsto 6$ \\
  o,p,q & $\mapsto 7$  & r,s,t & $\mapsto 8$ &  u,v,w & $\mapsto 9$ \\
        &              & x,y,z & $\mapsto 0$  
\end{tabular}
\caption{Example of stretched mapping}
\label{fig-stretch}
\end{center}
\end{figure}

\noindent Interestingly, the results for this mapping are very similar to the results of previous
method. In case of English dictionary, this method increases the coverings from 31\% to 48\% 
for the PIN length 4, and from 13\% to 16\% for the PIN length 5. For Slovak dictionary, we observed 
a slight increase from 7\% to 8\% for the PIN length 5. The covering percentage remained unchanged
for other two dictionaries.

\paragraph{Prefixes or suffixes} Another idea focuses on enlarging the space of PIN words. 
A straightforward approach is to use every dictionary word with length equal or greater than desired 
PIN length. The actual PIN words are then just the prefixes or the suffixes of these words. Both methods 
increase the covering of PINs with similar results. On the other hand, the distributions of
resulting PINs are skewed, because of common prefixes or suffixes 
(e.g. ``meth--'' or ``--ness'' for English). This yields a lower entropy of PINs generated by the suffix method
regardless of dictionary and PIN length, and by the prefix method applied to English and Slovak 
dictionaries regardless of PIN length. Hence, the methods do not present real improvement
over the standard approach from Section \ref{sec-stats}. Table \ref{tab-prefixes} shows the statistics 
for prefix method applied to English and French dictionaries.

\begin{table}[h]
\begin{center}
\begin{tabular}{lrrrr}
  & \multicolumn{2}{c}{English} 
  & \multicolumn{2}{c}{French} \\
  PIN length & \CA & \CB& \CA & \CB \\
  \noalign{\smallskip}\hline\noalign{\smallskip}
  Covering     &  39\% &  23\% & 29\%  &  11\% \\
  Entropy      & 10.56 & 13.80 & 10.39 & 12.43 \\
  \noalign{\smallskip}\hline
\end{tabular}
\caption{Results of prefix method for English and French dictionaries}
\label{tab-prefixes}
\end{center}
\end{table}

\paragraph{Combination} As another approach to improve the covering and the entropy of PINs,
we try to combine multiple methods. We use a combination of polyglot and prefixes methods with 
changed mappings of individual letters. The results can be viewed as some kind of ``upper bounds'' 
of what can be achieved by these and similar approaches. This combined approach reached 
the covering 86\% and 50\% for PIN lengths 4 and 5 respectively. The corresponding entropies 
of generated PINs were 11.43 and 13.78 respectively. These values are far from the
ideal values of uniformly generated PINs. Therefore, we try completely different
approaches in Section \ref{sec-hmm}.

\paragraph{Morphing PIN words with a digit} Unsurprisingly, better covering and entropy 
can be achieved by more complicated methods. Take for example the following procedure: choose a PIN
word and replace single letter at random position by some random digit. This way you can
morph the PIN word ``paper'' into ``p7per'' (or ``ppper'' if you want), ``pape0'', ``1aper'' etc. 
Then the morphed PIN word yield the PIN using standard mapping (a digit translates to itself). 
Hence you can get 40 or 50 PINs from a single PIN word of length 4 or 5, respectively. 
The results obtained by this methods, presented in Table \ref{tab-word1ex} for the PIN length 5, 
are even better than the results of combined approach. 

\begin{table}[h]
\begin{center}
\begin{tabular}{lrrrr}
  & English 
  & French
  & German
  & Slovak \\
  \noalign{\smallskip}\hline\noalign{\smallskip}
  Covering     &  67\% &  40\% & 44\% &   60\%  \\
  Entropy      & 15.22 & 14.47 & 14.79 & 15.12 \\
  \noalign{\smallskip}\hline
\end{tabular}
\caption{Statistics for morphing method (PIN length 5)}
\label{tab-word1ex}
\end{center}
\end{table}

On the other hand, this method can be too complex for a common user ((s)he must remember
the PIN word, the position and the value of the digit).

\section{Generating PIN Words}
\label{sec-hmm}

Experimenting with various natural ideas to improve the standard dictionary method in 
Section \ref{sec-improve} suggests that we should explore another possibilities. If ``distilling'' PIN
words from a dictionary is unsatisfactory, we will construct suitable PIN words from 
randomly and uniformly generated PINs, i.e. PINs with the highest entropy. By ``suitable PIN word'' 
we mean a word that is easy to remember. Certainly, it is very subjective what someone considers
as easy to memorize.

In order to allow all PINs, we must use a stretched mapping such as that one from the previous 
section. We use the mapping defined by Figure \ref{fig-stretch} for all experiments in this section.
However, the methods can be easily adapted to other mappings. 
The practical use of such mapping requires public availability of the mapping -- a user should 
not be required to memorize it. It can be achieved by printing the mapping on a numpad, displaying an 
image of such numpad to user as a part of an authorization dialog, etc.

\subsection{Hidden Markov Model}

We model translation of PIN words to PINs as a hidden Markov model \cite{Rab} 
with states being letters of the alphabet and digits of a PIN being observable 
data. The problem of finding suitable PIN word will be solved by the Viterbi algorithm.

More formally, let $Q$ be the set of states, and $V$ be the set of possible outputs. Let 
$\pi_i$ denotes the probability of starting in the state $i\in Q$. The probability of transition 
from the state $i$ to the state $j$ is denoted by $a_{i,j}$. Finally, let $b_{i}(k)$ be the probability 
of producing output $k\in V$ while being in the state $i\in Q$ .

We set $Q= \{\mbox{a},\mbox{b},\dots,\mbox{z}\}$, and $V = \{0,1,\dots, 9\}$.
The parameters of our hidden Markov model are instantiated according the particular dictionary.
The probability $\pi_i$ is computed as a fraction of dictionary words starting with character
$i\in Q$. The transition probability $a_{i,j}$, for $i,j\in Q$, corresponds to conditional 
probability that the next character is $j$ when the current character is $i$. The probability is 
calculated from all dictionary words. The output probabilities will reflect the deterministic
mapping of letters to digits, i.e. for $i\in Q$ and $k\in V$: $b_{i}(k) = 1$ if $i$ maps to 
$k$, otherwise $b_{i}(k)=0$.

A PIN can be generated randomly from uniform distribution and converted to the most 
suitable PIN word by finding the most probable sequence of states (letters) that produces
this PIN in our hidden Markov model. Since the sequence is based on the starting and
transition probabilities from a dictionary, the resulting PIN word should be easier
to remember than other PIN words for the same PIN. The problem of finding the most probable
sequence of states in hidden Markov model is well known, and the Viterbi algorithm can be used 
to solve it.

Table \ref{tab-hmm} shows examples of PIN words computed by the Viterbi algorithm for some randomly 
chosen PINs.

\begin{table}[h]
\begin{center}
\begin{tabular}{lllll}
  PIN & English & French & German & Slovak \\
  \noalign{\smallskip}\hline\noalign{\smallskip}
  1605    & \texttt{anyl} & \texttt{anyl} & \texttt{anzl} & \texttt{anyk} \\
  8407    & \texttt{thyp} & \texttt{thyp} & \texttt{sizo} & \texttt{sizo} \\
  2566    & \texttt{clmm} & \texttt{clmm} & \texttt{dlmm} & \texttt{dlnn} \\
  35130   & \texttt{flbex} & \texttt{elaex} & \texttt{flafz} & \texttt{ejbez} \\
  07588   & \texttt{zolst} & \texttt{zoltr} & \texttt{zoltr} & \texttt{zojst} \\
  94381   & \texttt{whera} & \texttt{viera} & \texttt{wiera} & \texttt{viera} \\
  \noalign{\smallskip}\hline
\end{tabular}
\caption{PIN words from hidden Markov model for some random PINs}
\label{tab-hmm}
\end{center}
\end{table}

The suitability of resulting PIN words is a subjective matter. However, the results are
not especially satisfying. The main problem is that there are PINs that cannot use vowels (i.e. `aeiouy') 
because of the particular mapping. This situation is unavoidable, the number of vowels is smaller 
than the number of digits. Therefore even using higher-order hidden Markov model will not
yield better results in such situations. As an alternative, we can generate a set of random PINs, and user 
chooses the one that (s)he found suitable for memorization. However, this makes the PINs 
not uniform anymore.

\subsection{PIN phrases}

Another approach is to use words instead of individual letters. The actual PIN word is a string
of the first letters from (dictionary) words in some sentence (we call it a PIN phrase). The sentence can be 
user-constructed or automatically generated. 
As a small experiment we implemented a simple PIN phrase generator from PINs using English 
words. For each digit we chose few nouns, verbs and adjectives with starting letter among the set
defined by the inverse of the stretched mapping. 
The selected words should allow for vivid visualizations to facilitate memorization. We defined 
a fixed sentence structure for the PIN length 5 -- adjective, noun, verb, adjective, noun. Some 
examples of generated PIN phrases for random PINs are shown in Table \ref{tab-phrases}.

\begin{table}[h]
\begin{center}
\begin{tabular}{ll}
  PIN & PIN phrase \\
  \noalign{\smallskip}\hline\noalign{\smallskip}
  05632 & \texttt{yellow lamb meets funny cat } \\
  19604 & \texttt{big witch makes yummy horse} \\
  90123 & \texttt{ugly zebra answers clean fly} \\
  38446 & \texttt{funny sun has happy mouse} \\
  \noalign{\smallskip}\hline
\end{tabular}
\caption{PIN phrases for some random PINs}
\label{tab-phrases}
\end{center}
\end{table}

A compromise between methods based on individual letters (such as the hidden Markov model method) 
and PIN phrases would be using just few words (e.g. a pair) and take two or three starting
letters from each word. For example ``\underline{cl}ear \underline{mn}emonic'' would represent 
PIN word ``clmn'', i.e. PIN 2566 using our stretched mapping.
Again, we can automate the process of translating a random PIN into such words. 

Generally, PIN phrases and their variants can be counted as mnemonic techniques. They can be further 
combined with or modified by other mnemonic systems, e.g. the major system. There is a vast number of 
possible modifications and customizations to these methods, and it is impossible to explore all of them here.

\section{Conclusion}

We analyzed several dictionary-based methods of choosing a PIN. Even though some individual's vocabulary is huge, the 
expected covering of the PIN space and the entropy of dictionary PINs are far from ideal. 
However, the results of the dictionary-based methods
are much better than the statistics of PINs usage observed in real-world applications. 
We also proposed two methods for constructing 
(hopefully) easy to remember PIN words for randomly chosen PINs. Their practical usability is a subjective 
matter, and therefore a suitable topic for further analysis and experiments.

Another interesting idea for subsequent research is the analysis of so-called geometric PINs, i.e.
PINs that are memorized according to the shape they create on the numpad.

\section*{Acknowledgments} The first author acknowledges support by the VEGA Research Grant 1/0781/11, 
the APVV Research Grant 0223-10, and the APVV support as part of the EUROCORES 
Programme EUROGIGA (project GREGAS, ESF-EC-0009-10) of the European Science 
Foundation. The second author acknowledges support by VEGA 1/0259/13.


\begin{thebibliography}{9}
\bibitem{PINstudy} N. Berry: \emph{PIN analysis}, DataGenetics, 2012. \\
  Available at www.datagenetics.com/blog/september32012/index.html (accessed 3 February 2013)
\bibitem{Vocab91} C.A. D'Anna,  E.B. Zechmeister, and J.W. Hall:
  \emph{Toward a meaningful definition of vocabulary size}, Journal of Reading Behavior, 
  23(1):109--122, 1991.
\bibitem{EMV}  EMVCo: \emph{EMV Integrated Circuit Card Specifications for Payment Systems}, 
  Version 4.3, 2011. \\
  Available at www.emvco.com/specifications.aspx (accessed 3 February 2013)
\bibitem{ISO}  International Organization for Standardization: 
  \emph{Financial services -- Personal Identification  Number (PIN) management and security},
  Part 1: Basic principles and requirements for PINs in card-based systems,  ISO 9564-1, 2011.
\bibitem{JL11} M. Jakobsson, D. Liu: \emph{Bootstrapping Mobile PINs Using Passwords}, 
  W2SP 2011: Web 2.0 Security and Privacy 2011.\\
  Available at w2spconf.com/2011/papers/mobilePIN.pdf (accessed 3 February 2013)
\bibitem{JL13} M. Jakobsson, D. Liu: \emph{Your Password is Your New PIN}, 
  SpringerBriefs in Computer Science 2013, pp 25-36, Springer, 2013.
\bibitem{PCI} PCI Security Standards Council: \emph{Payment Card Industry PIN Security Requirements}, 
  Version 1.0, 2011.\\
  Available at www.pcisecuritystandards.org/security\_standards (accessed 3 February 2013)
\bibitem{Pri} J. Pritchard: \emph{PIN Number Tips},  About.com Guide. \\
  Available at banking.about.com/od/securityandsafety/p/pinnumber.htm (accessed 3 February 2013)
\bibitem{Rab} L.R. Rabiner: \emph{A tutorial on hidden Markov models and selected
  applications in speech recognition}, Proceedings of the IEEE, Vol. 77, No. 2, 
  pp. 257--286, 1989. 
\bibitem{VISA} VISA: \emph{Issuer PIN Security Guidelines}, 2010.\\
  Available at usa.visa.com/merchants/risk\_management/cisp\_pin\_security.html (accessed 3 February 2013)
\bibitem{Vocab95} E.B. Zechmeister, A.M. Chronis, W.L. Cull, C.A. D'Anna and N.A. Healy: 
  \emph{Growth of a functionally important lexicon}, Journal of Reading Behavior, 27(2):201--212
  1995.
\end{thebibliography}
\end{document}